\documentclass[aps,prd,12pt,nofootinbib]{revtex4}

\usepackage{amsfonts}
\usepackage{amsmath,epsfig,bm}
\usepackage{graphicx}
\usepackage{color}

\usepackage[setpagesize=false]{hyperref}

\usepackage{tabularx}  %for table formatting 
\newcolumntype{s}{>{\centering\arraybackslash\hsize=.666\hsize}X} 
\newcolumntype{b}{>{\centering\arraybackslash\hsize=1.333\hsize}X} 

\usepackage{epsfig}

\newcommand{\be}{\begin{equation}}
\newcommand{\ee}{\end{equation}}
\newcommand{\ba}{\begin{eqnarray}}
\newcommand{\ea}{\end{eqnarray}}
\newcommand{\bd}{\begin{displaymath}}
\newcommand{\ed}{\end{displaymath}}

\def\thalf{{\textstyle{\frac{1}{2}}}}

\def\oneth{{\textstyle{\frac{1}{3}}}}
\def\twoth{{\textstyle{\frac{2}{3}}}}

\begin{document}

\title{Quasiparticle Theory of Transport Coefficients for Hadronic Matter at Finite Temperature and Baryon Density}

\author{M. Albright\footnote{albright@physics.umn.edu} and J. I. Kapusta\footnote{kapusta@physics.umn.edu}}
\affiliation{School of Physics \& Astronomy, University of Minnesota, Minneapolis, MN 55455,USA}
\
\date{August 11, 2015}

\begin{abstract}
We develop a flexible quasiparticle theory of transport coefficients of hot hadronic matter at finite baryon density.  We begin with a hadronic quasiparticle model which includes a scalar and a vector mean field. Quasiparticle energies and the mean fields depend on temperature and 
baryon chemical potential.  Starting with the quasiparticle dispersion relation, we derive the Boltzmann equation and use the Chapman-Enskog expansion to derive formulas for the shear and bulk viscosities and thermal conductivity.  We obtain both relaxation time approximation formulas
and more general integral equations.  Throughout the work, we explicitly enforce the
Landau-Lifshitz conditions of fit and ensure the theory is thermodynamically self-consistent.  The derived formulas should be useful for predicting
the transport coefficients of the hadronic phase of matter produced in heavy-ion collisions at the Relativistic Heavy Ion Collider (RHIC) and at other accelerators.
\end{abstract}

\maketitle

\parindent=20pt

\section{Introduction}

A central challenge in nuclear physics is elucidating the structure of the Quantum Chromodynamics (QCD) phase diagram.  Based on theoretical models, it is widely-believed that the phase diagram contains a line of first-order phase transition which ends at a point of second-order phase transition -- the critical point \cite{Stephanov,Mohanty2009}.  Despite a dedicated search for the critical point with the first beam energy scan at the 
Relativistic Heavy Ion Collider (RHIC) at Brookhaven National Lab, and possible hints of the critical point\cite{Lacey2015}, the location of the critical 
point remains a mystery.  From lattice QCD calculations \cite{Aoki2006,Aoki2009,Borsanyi2010,Bazavov2012,Borsanyi2014CrossoverTemp}, 
we know the transition from hadrons to quarks and gluons is an analytic crossover near temperature $T \approx 150$ MeV at zero baryon chemical potential $\mu_B$.  Hence, the critical point is generally thought to be located at $T < 160$ MeV and $\mu_B$ equal to several hundreds of MeV. 

A second, future beam energy scan at RHIC will search for the critical point with greatly increased statistics and upgraded detectors \cite{Odyniec2015}.  To maximize the discovery potential, experimental efforts must be accompanied by complementary improvements
in theoretical modeling of QCD matter at moderate temperatures and large baryon chemical potentials.  In previous papers, we investigated the equation of state at finite baryon chemical potential \cite{Albright2014,Albright2015}.  In this work, we derive new formulas to compute the shear and bulk viscosities and thermal conductivity of hot hadronic matter with $\mu_B >0$.  We employ a flexible, thermodynamically consistent framework of hadronic quasiparticles with medium-dependent quasiparticle masses and with a scalar and vector mean field.  This may be considered a natural extension of \cite{ChakrabortyKapusta2011} to include nonzero baryon chemical potential and the concomitant vector mean field.

Transport coefficients like the shear and bulk viscosities and thermal conductivity are especially interesting quantities to study for several reasons.
First, they are essential theoretical inputs for hydrodynamic simulations, which are critical tools for interpreting heavy-ion collision data.  In hydrodynamic simulations, the shear and bulk viscosities influence various observables, such as the elliptic flow coefficients $v_n$ and the hadron
transverse momentum ($p_T$) spectrum \cite{Song2008,Bozek2010,Dusling2012,NoronhaHostler2014}.  Furthermore, the temperature and chemical potential dependence of transport coefficients may reveal the location of phase transitions: in many physical systems, the shear viscosity is a minimum and the bulk viscosity a maximum at the phase transition \cite{Kapusta2010viscous}.  A third motivation is investigating the KSS lower bound \cite{Kovtun2005} on the shear viscosity to entropy density $\eta/s \ge 1/4\pi$ for strongly-coupled conformal theories and its implications for QCD.

In principle, the transport coefficients can be computed directly from QCD using the Kubo formulas \cite{Kubo1957}.  However, QCD is strongly coupled at energies accessible to heavy-ion collision experiments, complicating first-principles calculations.  There were some early attempts to employ lattice QCD \cite{Karsch1987,Sakai1998}, but even today it is challenging to achieve a large enough grid with a small enough grid
spacing to accurately compute transport coefficients.  Furthermore, lattice QCD simulations are currently very difficult at finite baryon chemical potential due to the well-known fermion sign problem.  Hence, many of the early works \cite{Hosoya1985,Gavin1985,Danielewicz1985,Prakash1993} 
computed transport coefficients of quark-gluon plasmas, or hadronic gases with a few species of particles, using the Boltzmann equation in the relaxation time approximation.  These early works did not include mean fields or medium-dependent masses.

Later on, Jeon \cite{Jeon1995} and Jeon and Yaffe \cite{Jeon1996} computed the shear and bulk viscosities of a hot, weakly-coupled scalar field theory using perturbation theory.  Amazingly, they showed that their complicated perturbative calculation of transport coefficients was 
reproduced by a simpler kinetic theory of quasiparticles with temperature-dependent masses and a scalar mean field. The same conclusion was found 
for hot, weakly-coupled QCD and QED \cite{Baym1990,Baym1991,Arnold2000,Arnold2003,Arnold2003a, Gagnon2007,Gagnon2007a,Gagnon2007b}.
This was also consistent with an earlier analysis of transport in a nucleon plus $\sigma$ meson system, which similarly found that renormalized quasiparticle masses were required \cite{Davis1991}.  Though astounding, this makes intuitive sense: kinetic theory is widely used to model 
non-equilibrium systems, and renormalized particle masses are ubiquitous in finite-temperature field theories.  (They are also present in Fermi liquid theory \cite{Baym2008}.)  Also, temperature- and chemical potential-dependent masses allow quasiparticle models to generate more realistic, 
non-ideal gas, equations of state \cite{Romatschke2012}.  Furthermore, as Gorenstein and Yang pointed out \cite{Gorenstein1995},
the scalar mean field is essential for maintaining thermodynamic self consistency when masses depend on temperature and/or chemical
potential.  Hence, it seems kinetic theories of quasiparticles with medium-dependent masses and mean fields are powerful theoretical tools, though thermodynamic consistency must be carefully maintained.
 
More recently, the conjecture of a lower bound on $\eta/s$ by Kovtun, Son, and Starinets from AdS/CFT \cite{Kovtun2005} ignited a flurry of additional work.  There were several more lattice calculations \cite{Nakamura2005,Sakai2005,Meyer2007,Meyer2009}.  There were also many studies with Boltzmann equations -- most of them without medium-dependent masses or mean fields.  Shear viscosity was computed for pion-nucleon gases at low temperatures and varying chemical potentials in  \cite{Chen2007,Itakura2008}.  Bulk viscosity of cool pion gases was computed using chiral perturbation theory in \cite{Itakura2008,Dobado2011}.  Shear viscosity in mixtures of hadrons with excluded volumes were calculated in \cite{Gorenstein2008,Noronha-Hostler2009,Noronha-Hostler2012}.

There were a few attempts to employ the more powerful quasiparticle models with medium-dependent masses to compute transport coefficients.  
In an early work, Sasaki and Redlich applied kinetic theory and the relaxation time approximation to a quasiparticle model to compute the bulk viscosity near a chiral phase transition \cite{Sasaki2009}.  Later, Chakraborty and one of us developed a comprehensive theory of shear and bulk viscosities in hadronic gases  \cite{ChakrabortyKapusta2011}.  That work included multiple hadron species with temperature-dependent masses and a scalar mean field in a thermodynamically self consistent way. They derived formulas for shear and bulk viscosity and provided both relaxation time approximation formulas and more general integral equations.  However, they did not include chemical potentials,  hence, thermal conductivity was not considered in that work.  Bluhm, K\"{a}mpfer, and Redlich used a similar quasiparticle formalism to study the shear and bulk viscosity of gluon matter in \cite{Bluhm2011} (also without chemical potentials).  Thus, a natural question is, how does the formalism of \cite{ChakrabortyKapusta2011} generalize to finite baryon chemical potential? Also, what is the formula for thermal conductivity?

Several papers have tried different ansatzes for generalized viscosity formulas (in the relaxation time approximation) when the baryon chemical potential is non-zero.  Chen, Liu, Song, and Wang calculated the shear and bulk viscosities of weakly-coupled quark gluon plasma at finite temperature and chemical potential in \cite{Chen2013} using a quasiparticle model with medium-dependent masses and a scalar mean field.  Khvorostukhin, Toneev, and Voskresensky compared three ansatzes for the generalized bulk viscosity formula \cite{Khvorostukhin2013} of a hadron gas with medium-dependent masses and a scalar mean field; see also \cite{Khvorostukhin2010,Khvorostukhin2011}.  Interestingly, Khvorostukhin's quasiparticle model also included a vector ($\omega$) mean  field\cite{Khvorostukhin2010,Khvorostukhin2013}; as is well known, they are important to account for repulsive forces in hadronic matter with large baryon densities.  This type of model is quite relevant for studying the moderate temperature hadronic matter formed in the beam energy scan at RHIC.  It is also relevant for experiments at the Super Proton Synchrotron Heavy Ion and Neutrino Experiment (SHINE) at CERN and at the future Facility for Antiproton and Ion Research (FAIR) at GSI.  Given the usefulness of this kind of model, it is desirable to put the results on a firmer theoretical foundation and (ideally) determine which of the 
ansatzes presented in \cite{Chen2013} and \cite{Khvorostukhin2013} are correct.

In this work, we present detailed derivations of the formulas for the shear and bulk viscosities and thermal conductivity of a gas of hadronic quasiparticles.  We include a scalar and a vector mean field, where the mean fields and the quasiparticle masses depend on temperature and
baryon chemical potential.  Generalization to multiple scalar and vector fields is straightforward but not included here for clarity of presentation.
Starting from the quasiparticle dispersion relation, we obtain the Boltzmann equation, and then use the Chapman-Enskog expansion
to derive formulas for the transport coefficients.  At each step we ensure that thermodynamic self-consistency is maintained, and we carefully enforce
the Landau-Lifshitz conditions of fit; we later show this is vital to obtaining the correct results.  We derive both relaxation time approximation formulas and more general integral equations.  Finally, we show that the formulas for shear and bulk viscosities are straightforward generalizations of
previous results \cite{ChakrabortyKapusta2011, Jeon1996} if one recalls that entropy per baryon is conserved in ideal hydrodynamics (neglecting viscous effects).  Classical statistics are used in the main text for ease of presentation, but results which include quantum statistics are presented in the appendix, albeit without detailed derivations.

\section{Quasiparticles}

In this section we discuss quasiparticle dispersion relations for baryons and mesons. In the simplest mean field approach all hadrons acquire effective masses in the medium.  In addition, baryons acquire effective chemical potentials.  We will focus attention on baryons since the inclusion of the baryon chemical potential is the new feature of this work compared to \cite{ChakrabortyKapusta2011}.

The piece of the Lagrangian involving baryons is
\be
{\cal L}_{\rm baryon} = \sum_j \bar{\psi}_j ( i \! \not\!\partial - m_j + g_{\sigma j} \sigma - g_{\omega j} \not\!\omega ) \psi_j \,.
\ee
Here $j$ refers to the species of baryon.  For simplicity of presentation we include only a generic scalar meson $\sigma$ and a generic vector meson $\omega$.  When evaluating the partition function there enters an additional term of the form $\mu_B  \bar{\psi}_j \gamma^0 \psi_j$, where $\mu_B$ is the baryon chemical potential.  Since we are using Dirac spinors both particles and antiparticles are included.  Particles have chemical potential $\mu_B$ while antiparticles have chemical potential $-\mu_B$.

For a uniform medium in thermal equilibrium the meson fields acquire space-time independent nonzero mean values denoted by $\bar{\sigma}$ and $\bar{\omega}^{\mu}$; in the rest frame of the medium the spatial part of the vector field vanishes on account of rotational symmetry, $\bar{\mbox{\boldmath $\omega$}}=0$, but in a general frame of reference it does not.  The dispersion relation for particles is
\be
E_j^+({\bf p}) = \sqrt{({\bf p} - g_{\omega j} \bar{\mbox{\boldmath $\omega$}})^2 + m_j^{*2}} + g_{\omega j} \bar{\omega}^0
\ee
and for antiparticles
\be
E_j^-({\bf p}) = \sqrt{({\bf p} + g_{\omega j} \bar{\mbox{\boldmath $\omega$}})^2 + m_j^{*2}} - g_{\omega j} \bar{\omega}^0 \,.
\ee
The kinetic momentum ${\bf p}^*$ is related to the canonical momentum ${\bf p}$ by
\be
{\bf p}_j^* = {\bf p} - g_{\omega j} \bar{\mbox{\boldmath $\omega$}}
\ee
for particles and by
\be
{\bf p}_j^* = {\bf p} + g_{\omega j} \bar{\mbox{\boldmath $\omega$}}
\ee
for antiparticles.  Particles and antiparticles have a common mass $m_j^*$.  In this mean field approach it is given by $m_j^* = m_j - g_{\sigma j} \bar{\sigma}$.

A more convenient way to think about the dispersion relations is to recognize a shift in both the mass and chemical potential of quasiparticles and anti-quasiparticles.  They both have energy
\be
E_j^{*\pm}({\bf p}^*) = \sqrt{{\bf p}^{*2}+ m_j^{*2}}
\ee
while their chemical potentials are opposite in sign
\be
\mu_j^{*\pm} = \pm (\mu_B - g_{\omega j} \bar{\omega}^0)
\ee 
as befits particles and antiparticles.

Mesons do not have a baryon chemical potential.  They could have chemical potentials for electric charge or strangeness, but we do not consider that possibility here for simplicity.  Hence their dispersion relations, in mean field approximation, are of the form
\be
E^*({\bf p}) = \sqrt{{\bf p}^{2}+ m^{*2}} \,.
\ee
Note that the kinetic and canonical momenta are the same for mesons.  The effective masses and effective chemical potentials can be found self-consistently once one fixes the Lagrangian.

In equilibrium the phase space density for a particle (or antiparticle) of type $a$ is given by
\be
f_a({\bf x},{\bf p}^*,t) = \frac{1}{{\rm e}^{(E_a^*-\mu_a^*)/T} - (-1)^{2s_a}} \,.
\ee
Here $s_a$ denotes the spin.  There are Fermi-Dirac and Bose-Einstein distributions.  Later on we will simplify our results by using classical statistics, although that approximation is not necessary.  Results including quantum statistics are given in the appendix.  Momentum space integration will be abbreviated as
\be
d\Gamma_a^* = (2s_a+1) \frac{d^3p_a^*}{(2\pi)^3}
\ee
indicating that the kinetic momentum is chosen as the independent variable, and the spin degeneracy is included.

\section{Boltzmann Equation}

The general form of the Boltzmann equation for the distribution function $f_a({\bf x},{\bf p}^*,t)$ is
\be
\frac{df_a}{dt}({\bf x},{\bf p}^*,t) = \frac{\partial f_a}{\partial t} + \frac{\partial f_a}{\partial x^i } \frac{d x^i}{dt} + \frac{\partial f_a}{\partial p^{*i}} \frac{d p^{*i}}{dt} =
{\cal C}_a \,.
\label{genform}
\ee
The right hand side is the collision term which will be discussed later.  Here we focus on the left hand side.  It involves the trajectory ${\bf x}(t)$ and ${\bf p}^*(t)$ between collisions.  This trajectory is in general not a straight line because the particle is moving in a mean field which can be space and time dependent.

The velocity is
\be
\frac{d x^i}{dt} = \frac{\partial E_a}{\partial p_a^i} = \frac{p^{*i}}{E_a^*} \,.
\ee
The relativistic version of Newton's Second Law is
\be
\frac{d p_a^i}{d t} = - \left( \frac{\partial E_a}{\partial x^i} \right)_p \,.
\ee
Note that it is $p$ that is held fixed, not $p^*$.  The right hand side is
\be
\left( \frac{\partial E_a}{\partial x^i} \right)_p = \frac{m_a^*}{E_a^*} \, \frac{\partial m_a^*}{\partial x^i} - g_{\omega a}  \frac{\partial \bar{\omega}^j}{\partial x^i}
 \frac{p^{*j}}{E_a^*} + g_{\omega a} \frac{\partial \bar{\omega}^0}{\partial x^i} \,.
\ee
The left hand side of Newton's Second Law can be written in terms of the kinetic momentum as 
\be
\frac{d p_a^i}{d t} = \frac{d p^{*i}}{d t} + g_{\omega a} \frac{d \bar{\omega}^i}{d t} = 
\frac{d p^{*i}}{d t} + g_{\omega a} \left( \frac{\partial \bar{\omega}^i}{\partial t} + \frac{p^{*j}}{E_a^*} \frac{\partial \bar{\omega}^i}{\partial x^j} \right) \,.
\ee

The time derivatives of ${\bf x}$ and ${\bf p}^*$ can now be replaced in Eq. (\ref{genform}) to put the Boltzmann equation in the form
\bd
\frac{df_a}{dt}({\bf x},{\bf p}^*,t) = \frac{\partial f_a}{\partial t} + \frac{p^{*i}}{E_a^*} \, \frac{\partial f_a}{\partial x^i }
\ed
\be
 - \frac{\partial f_a}{\partial p^{*i}}  \left\{ \frac{m_a^*}{E_a^*} \, \frac{\partial m_a^*}{\partial x^i} 
+ g_{\omega a} \left[ \frac{\partial \bar{\omega}^0}{\partial x^i} + \frac{\partial \bar{\omega}^i}{\partial t}
+ \frac{p^{*j}}{E_a^*} \left( \frac{\partial \bar{\omega}^i}{\partial x^j}  - \frac{\partial \bar{\omega}^j}{\partial x^i}  \right)\right]\right\} = {\cal C}_a \,.
\ee
This can be simplified by making use of the kinetic 4-momentum
\be
p_a^{*\mu} = \left( E_a^*,{\bf p}^* \right)
\ee
and  the field strength tensor
\be
\omega^{\alpha \beta} \equiv \partial^{\alpha} \omega^{\beta} - \partial^{\beta} \omega^{\alpha} \,.
\ee
The final form is
\be
\frac{df_a}{dt}({\bf x},{\bf p}^*,t) = \frac{p^{*\mu}}{E_a^*} \partial_{\mu} f_a - \left[ \frac{m_a^*}{E_a^*} \, \frac{\partial m_a^*}{\partial x^i} 
+ g_{\omega a}  \frac{p_{\mu}^*}{E_a^*} \,  \bar{\omega}^{\mu i} \right] \frac{\partial f_a}{\partial p^{*i}} = {\cal C}_a \,.
\label{finform}
\ee

\section{Energy-Momentum Tensor and Baryon Current}

In this section we present the structure of the energy-momentum tensor $T^{\mu\nu}$ and of the baryon current $J_B^{\mu}$.  In terms of temperature, chemical potential, and flow velocity $u^{\mu}$ they are
\be
T^{\mu\nu} = -P g^{\mu\nu} + w u^{\mu} u^{\nu} + \Delta T^{\mu\nu}
\label{Tuv}
\ee
and
\be
J_B^{\mu} = n_B u^{\mu} + \Delta J_B^{\mu}
\ee
where $P(T,\mu_B)$ is the pressure, $s = \partial P/\partial T$ is the entropy density, $n_B = \partial P/\partial \mu_B$ is the baryon density, $\epsilon = -P + Ts + \mu_B n_B$ is the energy density, and $w = \epsilon + P$ is the enthalpy density.  In the Landau-Lifshitz approach, which we use, $u^{\mu}$ is the velocity of energy transport.  The $\Delta T^{\mu\nu}$ and $\Delta J_B^{\mu}$ are dissipative parts given by
\be
\Delta T^{\mu\nu} = \eta \left( D^{\mu} u^{\nu} + D^{\nu} u^{\mu} + \twoth \Delta^{\mu\nu} \partial_{\rho} u^{\rho} \right) - \zeta \Delta^{\mu\nu} \partial_{\rho} u^{\rho}
\label{DTuv}
\ee
and 
\be
\Delta J_B^{\mu} = \lambda \left( \frac{n_B T}{w}\right)^2 D^{\mu} \left( \frac{\mu_B}{T} \right) \,.
\label{DJu}
\ee
Here $\eta$, $\zeta$ and $\lambda$ are the shear viscosity, bulk viscosity, and thermal conductivity, respectively.  The other symbols are
\be
D = u^{\rho} \partial_{\rho} \, , 
\ee
\be
D^{\mu} = \partial^{\mu} - u^{\mu}D \, ,
\ee
\be
\Delta^{\mu\nu} = u^{\mu}u^{\nu} - g^{\mu\nu} \,.
\ee
Our metric is $(+,-,-,-)$.  Additionally, the entropy current is
\be
s^{\mu} = s u^{\mu} - \frac{\mu_B}{T} \Delta J_B^{\mu} \,.
\ee

Now we need to express $T^{\mu\nu}$ and $J_B^{\mu}$ in terms of the quasiparticles and mean fields.  One expression for the former is
\be
T^{\mu\nu} = \sum_a \int d\Gamma_a^* \frac{p_a^{*\mu} p_a^{*\nu}}{E_a^*} f_a  + g^{\mu\nu} U(\bar{\sigma}, \bar{\omega}^{\rho} \bar{\omega}_{\rho})
+ m_{\omega}^2 \bar{\omega}^{\mu} \bar{\omega}^{\nu} \,.
\label{Tuvf*}
\ee
The first term is familiar as the kinetic contribution.  The second term is the usual meson field potential energy; it includes the mass terms $\thalf m_{\sigma}^2 \bar{\sigma}^2$ and 
$-\thalf m_{\omega}^2 \bar{\omega}^{\rho} \bar{\omega}_{\rho}$, plus any interaction terms which are more than two powers of the fields.  Note that kinetic terms for the mean meson fields are not included because they are second order in space-time gradients and are not included in first order viscous fluid dynamics.  The last term is not obviously of the form of Eq. (\ref{Tuv}).  However, when one remembers that $T^{0i}$ is the energy flux in the direction $i$, and that $E_a$ is the complete quasiparticle energy and not $E_a^*$, then one would write
\be
T^{\mu\nu} = \sum_a \int d\Gamma_a^* \frac{p_a^{\mu} p_a^{*\nu}}{E_a^*} f_a +
g^{\mu\nu} U(\bar{\sigma}, \bar{\omega}^{\rho} \bar{\omega}_{\rho}) \,.
\label{Tuvf}
\ee
Using $p_a^{\mu} = p_a^{*\mu} + g_{\omega a} \bar{\omega}^{\mu}$ we get
\be
T^{\mu\nu} = \sum_a \int d\Gamma_a^* \frac{p_a^{*\mu} p_a^{*\nu}}{E_a^*} f_a  + g^{\mu\nu} U(\bar{\sigma}, \bar{\omega}^{\rho} \bar{\omega}_{\rho})
+ \bar{\omega}^{\mu} \sum_a g_{\omega a} \int d\Gamma_a^* \frac{p_a^{*\nu}}{E_a^*} f_a \,.
\ee
The vector mean field is determined by its equation of motion.  Assuming an interaction only with the baryons (this assumption is easily relaxed) it is
\be
\left( \partial^2 + m_{\omega}^2 \right) \bar{\omega}^{\nu} = \sum_j g_{\omega j} \langle \bar{\psi}_j \gamma^{\nu} \psi_j \rangle
\ee
where the averaging refers to the quasiparticle distribution.  Recognizing that the summation index $j$ refers to both baryons and antibaryons, and dropping the d'Alembertian because of first order viscous fluid dynamics, we have
\be
m_{\omega}^2 \bar{\omega}^{\nu} = \sum_a g_{\omega a} \int d\Gamma_a^* \frac{p_a^{*\nu}}{E_a^*} f_a \,.
\ee
(We remind the reader that the coupling $g_{\omega a}$ is opposite in sign for baryons and antibaryons.)  Hence Eqs. (\ref{Tuvf*}) and (\ref{Tuvf}) are the same.

In a similar way the scalar mean field is determined by its equation of motion.  This turns out to be
\be
\frac{\partial U(\bar{\sigma}, \bar{\omega}^{\rho} \bar{\omega}_{\rho})}{\partial \bar{\sigma}} = \sum_a g_{\sigma a} \int d\Gamma_a^* \frac{m_a^*}{E_a^*} f_a \,.
\ee
The coupling to scalar mesons of baryons and antibaryons has the same sign, unlike the coupling to vector mesons.

The structure of the baryon current is readily deduced to be
\be
J_B^{\mu} = \sum_a b_a \int d\Gamma_a^* \frac{p_a^{*\mu}}{E_a^*} f_a
\label{Jf} 
\ee
where $b_a$ denotes the baryon number of $a$.

It can be shown that energy and momentum are conserved, namely
\be
\partial_{\mu} T^{\mu\nu} = 0
\ee
and so is baryon number
\be
\partial_{\mu} J_B^{\mu} = 0 \,.
\ee
These conservation laws follow from the requirement that
\be
\sum_a \int d\Gamma_a^* \chi_a {\cal C}_a = 0 \, .
\ee
The $\chi_a$ represents the contribution from quasiparticle $a$ to any conserved quantity, such as energy, momentum, or baryon number.  The calculations are straightforward but very lengthy and tedious.  We have performed them but they are not reproduced here.  It is also straightforward, and much less tedious, to show that the mean field equation of state follows from the above expressions for $T^{\mu\nu}$ and $J_B^{\mu}$ when the system is uniform, time independent, and in thermal and chemical equilibrium.

\section{Departures from Equilibrium of the Quasiparticle Distribution Function}
\label{DeptartFromEquilib}

To first order in departures from equilibrium, we can express the quasiparticle distribution function as
\be
f_a = f_a^{\rm eq} \left( 1 + \phi_a \right)
\ee
where $f_a^{\rm eq}$ is the distribution function in thermal and chemical equilibrium.  The nonequilibrium part $\phi_a$ leads to the nonequilibrium contributions
$\Delta T^{\mu\nu}$ and $\Delta J_B^\mu$, so $\phi_a$ must contain the same space-time gradients as found in them. Therefore, $\phi_a$ must have the form 
\be
\phi_a = -A_a \partial_\rho u^\rho - B_a p_a^{\nu}  D_\nu \left( \frac{\mu_B}{T} \right)   + C_a p_a^{\mu} p_a^{\nu} 
\left( D_\mu u_\nu + D_\nu u_\mu + \twoth \Delta_{\mu\nu}\partial_\rho u^\rho \right) \, .
\label{eq:derivingphi:phi}
\ee
The functions $A_a$, $B_a$ and $C_a$ only depend on momentum $p$ while $u^{\mu}$ only depends on space-time coordinate $x$.  

The departure from equilibrium of the quasiparticle distributions can be used to compute the departure from equilibrium of the energy-momentum tensor.  It is convenient to work in the local rest frame.  The variation of the space-space part of expression (\ref{Tuvf*}) is
\be
\delta T^{ij} = \sum_a \int d\Gamma_a^* \frac{p_a^{*i} p_a^{*j}}{E_a^*} \left( \delta f_a - f_a^{\rm eq} \frac{\delta E_a^*}{E_a^*} \right) + g^{ij} \delta U \,.
\ee
To obtain the variation in the mean field potential we start with the expression for the pressure $P(T,\mu_B) = P_0 - U$.  Here $P_0$ is the kinetic contribution to the pressure from the quasiparticles.  The entropy density is obtained from $s = \partial P(T,\mu_B)/\partial T$.  This has three contributions: the first is from $s_0$ which is the same functional form as for particles with $T$- and $\mu_B$-independent energies, the second is from the variation of the quasiparticle energies due to variations in $T$ and $\mu_B$, and finally there is the contribution $-\partial U/\partial T$ at fixed $\mu_B$.  The mean field carries no entropy, therefore the second and third terms must cancel.  Using classical statistics for simplicity we have
\be
P_0 = T  \sum_a \int d\Gamma_a^* f_a^{\rm eq}
\ee
and thus
\be
\frac{\partial U}{\partial T} = - \sum_a \int d\Gamma_a^* \left(\frac{\partial E_a}{\partial T}\right)_{\mu_B} f_a^{\rm eq} \,.
\ee
The same argument applies to differentiation with respect to $\mu_B$, which gives the baryon density.  The mean field carries no baryon number, so similarly
\be
\frac{\partial U}{\partial \mu_B} = - \sum_a \int d\Gamma_a^* \left(\frac{\partial E_a}{\partial \mu_B}\right)_T f_a^{\rm eq} \,.
\ee
Hence
\be
\delta U = - \sum_a \int d\Gamma_a^* \delta E_a f_a^{\rm eq}
\ee
where $E_a = E_a^* + g_{\omega a} \bar{\omega}^0$ and
\be
\delta E_a = \frac{m_a^*}{E_a^*} \delta m_a^* + g_{\omega a} \delta \bar{\omega}^0 \, .
\ee 

Now we come to the deviation in the quasiparticle distribution function.  The $f_a$ in general will have departures from the equilibrium form, but it can also change because the quasiparticle energy departs from its equilibrium value.  Let us denote $E_a^0$ the equilibrium value and $E_a$ the total nonequilibrium energy; it is the latter which is conserved in the particle collisions.  Similarly, we denote $T^0$ and $\mu_B^0$ the equilibrium values.  Then we write 
\ba
f_a(E_a, T, \mu_B) &=& f_a^{\rm eq}(E_a^0, T^0, \mu_B^0) + \delta f_a \, , \nonumber \\
f_a(E_a, T, \mu_B) &=& f_a^{\rm eq}(E_a, T^0, \mu_B^0) + \delta \tilde{f}_a \,.
\ea
The deviations are related to each other by
\be
\delta f_a = \delta \tilde{f}_a + \left( \frac{\partial f_a^{\rm eq}}{\partial E_a} \right)_{T^0 ,\, \mu_B^0} \delta E_a
= \delta \tilde{f}_a - \frac{\delta E_a}{T} f_a^{\rm eq}
\label{ftildef}
\ee
where the second equality follows when using classical statistics.

It is always the $\delta \tilde{f}_a$ which determine the transport coefficients.  Therefore we express $\delta T^{ij}$ in terms of $\delta \tilde{f}_a$ instead of $\delta f_a$.
\ba
\delta T^{ij} &=& \sum_a \int d\Gamma_a^* \frac{p_a^{*i} p_a^{*j}}{E_a^*} \delta \tilde{f}_a 
- \sum_a \int d\Gamma_a^* \frac{p_a^{*i} p_a^{*j}}{E_a^*} \left( \frac{\delta E_a}{T} 
+ \frac{\delta E_a^*}{E_a^*} \right) f_a^{\rm eq} \nonumber \\
&+& \delta^{ij} \sum_a \int d\Gamma_a^* \delta E_a f_a^{\rm eq}
\ea
The integrand of the second term depends only on the magnitude of ${\bf p}_a^*$, apart from the factor $p_a^{*i} p_a^{*j}$.  Therefore, one may effectively make the replacement $p_a^{*i} p_a^{*j} \rightarrow \oneth |{\bf p}_a^*|^2 \delta^{ij}$.  Then the terms not involving $\delta \tilde{f}_a$ all have a factor of $\delta^{ij}$.  They can be written as a sum of
\bd
\frac{\delta \bar{\omega}^0}{T} \sum_a g_{\omega a} \int d\Gamma_a^* \left( T - \frac{|{\bf p}_a^*|^2}{3 E_a^*} \right) f_a^{\rm eq}
\ed
and
\bd
\sum_a \delta m_a^*\int d\Gamma_a^* \frac{m_a^*}{E_a^*}  \left( 1 - \frac{|{\bf p}_a^*|^2}{3 T E_a^*} - \frac{|{\bf p}_a^*|^2}{3 E_a^{*2}} \right) f_a^{\rm eq} \,.
\ed
It can be shown that both of these integrate to zero (using classical statistics).  Hence we find
\be
\delta T^{ij} = \sum_a \int d\Gamma_a^* \frac{p_a^{*i} p_a^{*j}}{E_a^*} \delta \tilde{f}_a
\label{dTij}
\ee
as our final result.

The variation in the time-time component of the energy-momentum tensor, starting with either Eq. (\ref{Tuvf*}) or (\ref{Tuvf}), is 
\be
\delta T^{00} = \sum_a \int d\Gamma_a^* E_a \delta f_a \,.
\label{energyvar}
\ee
We use Eq. (\ref{ftildef}) for $\delta f_a$.  The variation of the local energy $E_a$ 
\be
\delta E_a =  \frac{\delta m_a^{*2}}{2 E_a^*} + g_{\omega a} \delta \bar{\omega}^0 
\ee
can be expressed in terms of the variations in temperature and chemical potential
\ba
\delta m_a^{*2} &=& \left( \frac{\partial m_a^{*2}}{\partial T} \right)_{\!\! \mu_B} \delta T 
+  \left( \frac{\partial m_a^{*2}}{\partial \mu_B} \right)_{\!\! T} \delta \mu_B \, , \\
\delta \bar{\omega}^0 &=& \left( \frac{\partial \bar{\omega}^0}{\partial T} \right)_{\!\! \mu_B} \delta T 
+  \left( \frac{\partial \bar{\omega}^0}{\partial \mu_B} \right)_{\!\! T} \delta \mu_B \,.
\ea
The variations $\delta T$ and $\delta \mu_B$ are not independent.  They are related by the hydrodynamic flow of the matter which to this order occurs at constant entropy per baryon $\sigma = s/n_B$.  Dissipation should not be included since it would lead to second-order effects which are consistently neglected in first order viscous fluid dynamics.  The relation can be expressed in various ways, including these:
\be
\left( \frac{\partial \mu_B}{\partial T} \right)_{\!\! \sigma} = \frac{\mu_B}{T} \frac{v_s^2}{v_n^2} 
= \frac{1}{T} \left[ \mu_B + \frac{1}{v_n^2} \left( \frac{\partial P}{\partial n_B} \right)_{\!\! \epsilon} \, \right]
= \frac{ \chi_{TT} - \sigma \chi_{\mu T}}{\sigma \chi_{\mu\mu} - \chi_{\mu T}} \, .
\label{isentropic}
\ee
Here $v_x^2 = (\partial P/\partial \epsilon)_x$ is the speed of sound at constant $x$.  It is easily shown that
\ba
v_n^2 &=& \frac{s \chi_{\mu\mu} - n_B \chi_{\mu T}}{T(\chi_{TT} \chi_{\mu\mu} - \chi^2_{\mu T})} \, , \nonumber \\
v_s^2 &=& \frac{n_B \chi_{TT} - s \chi_{\mu T}}{\mu_B (\chi_{TT} \chi_{\mu\mu} - \chi^2_{\mu T})} \, , \nonumber \\
v_{\sigma}^2 &=& \frac{v_n^2 T s + v_s^2 \mu n_B}{w} \, ,
\label{speeds}
\ea
relationships that are independent of the specific equation of state.  Of course waves do not physically propagate at constant $n$ or $s$, only at constant $\sigma$, but these definitions are useful for various intermediate steps in various applications.  The other symbol represents the susceptibilities
\be
\chi_{xy} = \frac{\partial^2 P(T,\mu)}{\partial x \partial y} \,.
\ee
Rather than thinking of $m_a^*$ and $\bar{\omega}^0$ as functions of $T$ and $\mu_B$ we can think of them as functions of $T$ and $\sigma$.  Then
\ba
\delta m_a^{*2} &=& \left( \frac{\partial m_a^{*2}}{\partial T} \right)_{\!\! \sigma} \delta T \, , \\
\delta \bar{\omega}^0 &=& \left( \frac{\partial \bar{\omega}^0}{\partial T} \right)_{\!\! \sigma} \delta T  \,.
\ea

Next, we need to relate the variations in $T$ and $\mu_B$ to the variation $\delta \tilde{f}_a$.  The latter variation is done at fixed $E_a$ and is
\be
\delta \tilde{f}_a = f_a^{\rm eq} \left[ E_a - \mu_a + T \left( \frac{\partial \mu_a}{\partial T} \right)_{\!\! \sigma} \right]
\frac{\delta T}{T^2} \, .
\label{dtildef}
\ee
(Recall that $\mu_a = b_a \mu_B$.)  The term from Eq. (\ref{ftildef}) which needs to be rewritten is
\ba
\frac{\delta E_a}{T} f_a^{\rm eq} &=& \frac{1}{E_a^*} \left[ 
\frac{T^2 \left( \partial m_a^{*2} / \partial T^2 \right)_{\sigma} + g_{\omega a} T \left( \partial \bar{\omega}^0 / \partial T \right)_{\sigma} E_a^* }
{E_a - \mu_a + T \left( \partial \mu_a / \partial T \right)_{\sigma} } \right] \delta \tilde{f}_a \nonumber \\
&=& \left[ \frac{ T( \partial E_a/\partial T)_{\sigma}}{E_a - \mu_a + T \left( \partial \mu_a / \partial T \right)_{\sigma} } \right] \delta \tilde{f}_a \,.
\ea
We reiterate that the temperature derivative of a function $F$ depending on $T$ and $\mu_B$, taken at fixed entropy per baryon, is
\be
\left( \frac{\partial F}{\partial T} \right)_{\!\! \sigma} = \left( \frac{\partial F}{\partial T} \right)_{\!\! \mu_B} 
+ \left( \frac{\partial F}{\partial \mu_B} \right)_{\!\! T} \left( \frac{\partial \mu_B}{\partial T} \right)_{\!\! \sigma}
=  \left( \frac{\partial F}{\partial T} \right)_{\!\! \mu_B} 
+ \frac{\mu_B}{T} \frac{v_s^2}{v_n^2} \left( \frac{\partial F}{\partial \mu_B} \right)_{\!\! T} \, .
\label{sigmaderivative}
\ee
The final expression is therefore
\be
\delta T^{00} = \sum_a \int d\Gamma_a^* E_a
 \left\{ 1  -   \frac{ T( \partial E_a/\partial T)_{\sigma}}{E_a - \mu_a + T \left( \partial \mu_a / \partial T \right)_{\sigma} }\right\} \delta \tilde{f}_a \,. 
\ee
When the baryon density goes to zero this reduces to the formula known in the literature.

The time-space component has the very natural form
\be
\delta T^{0j} = \sum_a \int d\Gamma_a^* \frac{p_a^{*j}}{E_a^*} E_a \delta f_a \,.
\ee
To express this in terms of $\delta \tilde{f}_a$, we note that the last term on the right-hand side of Eq. (\ref{ftildef}) is spherically symmetric in momentum space and therefore that term integrates to zero.  This is not true of the other term because the deviation $\phi_a$ does have terms that depend on the direction of the momentum.  Therefore
\be
\delta T^{0j} = \sum_a \int d\Gamma_a^* \frac{p_a^{*j}}{E_a^*} E_a \delta \tilde{f}_a \,.
\label{dspace-time} 
\ee

Lastly we need the variations in the baryon current.  The steps are by now very familiar.  The results are 
\be
\delta J_B^0 = \sum_a b_a \int d\Gamma_a^* 
\left\{ 1  -   \frac{ T( \partial E_a/\partial T)_{\sigma}}{E_a - \mu_a + T \left( \partial \mu_a / \partial T \right)_{\sigma} }  \right\} \delta \tilde{f}_a 
\ee
and
\be
\delta J_B^i = \sum_a b_a \int d\Gamma_a^* \frac{p_a^{*i}}{E_a^*} \delta \tilde{f}_a \,.
\ee

\section{General Formulas for the Transport Coefficients}

Suppose that we know the scalars $A_a$, $B_a$, and $C_a$ in Eq. (\ref{eq:derivingphi:phi}) as functions of the magnitude of the momentum ${\bf p}_a^*$.  Then in the local rest frame we should equate the hydrodynamic expression $\Delta T^{ij}$ from Eq. (\ref{DTuv}) with the quasiparticle expression $\delta T^{ij}$ from Eq. (\ref{dTij}), the latter being
\ba
\delta T^{ij} &=& \sum_a \int d\Gamma_a^* \frac{p_a^{*i} p_a^{*j}}{E_a^*} \Big[  -A_a \partial_\rho u^\rho - B_a p_a^{\nu}  D_\nu \left( \frac{\mu_B}{T} \right) \nonumber \\
&+& C_a p_a^{\mu} p_a^{\nu} \left( D_\mu u_\nu + D_\nu u_\mu + \twoth \Delta_{\mu\nu}\partial_\rho u^\rho \right) \Big] f_a^{\rm eq} \,.
\ea
The $B_a$ integrates to zero by symmetry.  In the local rest frame the derivative $\partial_k u_0 = 0$, so the the summation over $\mu$ and $\nu$ is a sum over spatial indices $kl$ only.  
In the $A_a$ term we can use
\bd
p_a^{*i} p_a^{*j} \rightarrow \oneth |{\bf p}_a^*|^2 \delta_{ij}
\ed
and in the $C_a$ term we can use
\bd
p_a^{*i} p_a^{*j}p_a^{*k} p_a^{*l} \rightarrow {\textstyle{\frac{1}{15}}} |{\bf p}_a^*|^4 ( \delta_{ij} \delta_{kl} + \delta_{ik} \delta_{jl} + \delta_{il} \delta_{jk} )
\ed
because in the local rest frame ${\bf p} = {\bf p}^*$.  Equating the tensorial structures then gives us the shear viscosity
\be
\eta = \frac{2}{15} \sum_a \int d\Gamma_a^* \frac{|{\bf p}_a^*|^4}{E_a^*} f_a^{\rm eq} C_a
\label{eta}
\ee
and the bulk viscosity
\be
\zeta = \frac{1}{3} \sum_a \int d\Gamma_a^* \frac{|{\bf p}_a^*|^2}{E_a^*} f_a^{\rm eq} A_a \,.
\label{zeta}
\ee

For the baryon current we compare the $\Delta J_B^i$ from Eq. (\ref{DJu}) with the dissipative part of Eq. (\ref{Jf}) in the local rest frame.  The latter is
\be
 \delta J_B^i = \sum_a b_a \int d\Gamma_a^* \frac{p_a^{*i}}{E_a^*} \left[ - B_a p_a^{\nu}  D_\nu \left( \frac{\mu_B}{T} \right) \right] f_a^{\rm eq} \,.
\ee
Obviously the $A_a$ and $C_a$ terms integrate to zero on account of symmetry.  After some manipulation this results in an expression for the thermal conductivity
\be
\lambda = \frac{1}{3} \left( \frac{w}{n_B T} \right)^2 \sum_a b_a \int d\Gamma_a^* \frac{|{\bf p}_a^*|^2}{E_a^*} f_a^{\rm eq} B_a \,.
\label{lambda}
\ee

To solve for the functions $A_a$, $B_a$, and $C_a$ we turn to the Chapman-Enskog method.  This entails expanding both sides of the Boltzmann equation (\ref{finform}) to first order in the $\phi_a$.  It leads to integral equations which in general must be solved numerically.

Here we follow the notation of \cite{ChakrabortyKapusta2011}.  Including 2-to-2, 2-to-1 and 1-to-2 processes, and using classical statistics (these restrictions are easily relaxed) the collision integral is 
\ba
 \mathcal{C}_a 
 &=& \sum_{bcd} \frac{1}{1+\delta_{ab}}
 \int d\Gamma_b^* \, d\Gamma_c^* \, d\Gamma_d^* \,
 W(a,b|c,d) \{f_c f_d - f_a f_b\} \nonumber \\
 &&
 + \sum_{cd} \int d\Gamma_c^* \, d\Gamma_d^* \, W(a|c,d) \{f_c f_d - f_a\}
 \nonumber \\
 &&
 + \sum_{bc} \int d\Gamma_b^* \, d\Gamma_c^* \, W(c|a,b) \{f_c - f_a f_b\} \,.
\label{eq:collisionintegral:collision_full1}
\ea
The $W$ are given as
\be
 W(a,b|c,d) = \frac{(2\pi)^4 \delta^4(p_a + p_b - p_c - p_d ) }
 {2E_a^* 2E_b^* 2E_c^* 2E_d^*} 
 | \overline{\mathcal{M}(a,b|c,d)} |^2   \,
\label{eq:collisionintegral:W2to2}
\ee
and
\be
 W(a|c,d) = \frac{(2\pi)^4 \delta^4(p_a - p_c - p_d ) }
 {2E_a^* 2E_c^* 2E_d^*} 
 | \overline{ \mathcal{M}(a|c,d) } |^2  \,.
 \label{eq:collisionintegral:W1to2}
\ee
The use of $E_a^*$ instead of $E_a$ in the denominators ensures that the phase space integration is Lorentz covariant.  Also note that, following Larionov \cite{Larionov2007}, we use dimensionless matrix elements $\mathcal{M}$
averaged over spin in both initial and final states.  This is necessary to balance the degeneracy factors in the $d\Gamma_a^*$.  We use chemical equilibrium (for example, $a + b \leftrightarrow c + d$ gives $f_a^{\rm eq} f_b^{\rm eq} = f_c^{\rm eq} f_d^{\rm eq}$.)  Then the collision integral becomes
\ba
 \mathcal{C}_a 
 &=& f_a^{\rm eq} \sum_{bcd} \frac{1}{1+\delta_{ab}}
 \int d\Gamma_b^* \, d\Gamma_c^* \, d\Gamma_d^* \,
 f_b^{\rm eq} W(a,b|c,d) \left[ \phi_c + \phi_d - \phi_a - \phi_b \right] \nonumber \\
 &+& f_a^{\rm eq} \sum_{cd} \int d\Gamma_c^* \, d\Gamma_d^* \, W(a|c,d) 
 \left[ \phi_c + \phi_d - \phi_a \right]
 \nonumber \\
 &+& f_a^{\rm eq} \sum_{bc} \int d\Gamma_b^* \, d\Gamma_c^* \,
 f_b^{\rm eq} W(c|a,b) \left[ \phi_c - \phi_a - \phi_b \right] \,.
 \label{eq:collisionintegral:collision_full2}
\ea
This constitutes the right-hand-side of the Boltzmann equation.

The left-hand side of the Boltzmann equation (\ref{finform}) is computed using the local equilibrium form of the distribution function
\be
f_a^{\rm eq}(x,{\bf p}^*) = \exp\left[ -\frac{u_{\alpha}(x) p_a^{\alpha}}{T(x)}\right]  \exp\left[ \frac{\mu_a(x)}{T(x)}\right]
= \exp\left[ -\frac{u_{\alpha}(x) p_a^{*\alpha}}{T(x)}\right]  \exp\left[ \frac{\mu_a^*(x)}{T(x)}\right] \,.
\ee
Here the flow velocity, temperature and chemical potential all depend on $x$.  Although not explicitly indicated, $p_a^{\alpha}$ depends on $x$ via the dependence of $m_a^*$ and $\bar{\omega}^{\alpha}$ on $x$, while $E_a^*$ depends on $x$ via $m_a^*$ only.  The left-hand side must be expressed in terms of the same space-time gradients as $\phi_a$, namely $\partial_\rho u^\rho$, $D_\nu \left( \mu_B/T \right)$, and $\left( D_\mu u_\nu + D_\nu u_\mu + \twoth \Delta_{\mu\nu}\partial_\rho u^\rho \right)$.  The calculation is long and tedious.  Space-time derivatives of $T$ and $\mu_B$ are expressed in terms of the relevant tensor structures by using the perfect fluid equations for conservation of energy, momentum and baryon number.  Some useful intermediate results are
\ba
D T &=& - v_n^2 T \, \partial_\rho u^\rho \nonumber \, , \\
D \mu_B &=& - v_s^2 \mu_B \, \partial_\rho u^\rho \,.
\ea
One form of the left-hand side (in the local rest frame) is
\ba
\frac{d f_a^{\rm eq}}{dt} &=&  f_a^{\rm eq} \left[  \frac{ |{\bf p}_a^*|^2 }{3 T E_a^*} 
 +  v_n^2 T \frac{\partial}{\partial T} \left(\frac{ E_a - \mu_a}{T} \right)_{\!\! \sigma} \right] \partial_\rho u^\rho \nonumber \\
 &+& f_a^{\rm eq} \left( b_a - \frac{n_B E_a}{w} \right)
 \frac{p_a^{\mu}}{E_a^*} D_\mu \left( \frac{\mu_B}{T} \right)  \nonumber \\
 & -& f_a^{\rm eq} \, \frac{p_a^{\mu} p_a^{\nu}}{2 T E_a^*} 
 \left( D_\mu u_\nu +  D_\nu u_\mu  + \frac{2}{3} \Delta_{\mu\nu} \partial_\rho u^\rho \right) \,.
\label{eq:chapman:BTE_LHS6}
\ea
Now $E_a - \mu_a$ in the first line could be replaced by $E_a^* - \mu_a^*$, and $E_a$ in the second line could be replaced by $E_a^* + g_{\omega a} \bar{\omega}^0$.  With a little manipulation this can be shown to be equivalent to Sasaki and Redlich who, however, did not include a vector field nor the $D_\mu ( \mu_B/T )$ term.  Another form is to write out the derivatives in the first line explicitly.  This results in
\ba
\frac{d f_a^{\rm eq}}{dt} &=&  f_a^{\rm eq} \, \frac{1}{3 T E_a^*} \left\{ |{\bf p}_a^*|^2 
 - 3 v_n^2 \left[ E_a^{*2} - T^2 \left( \frac{\partial m_a^{*2}}{\partial T^2} \right)_{\!\! \sigma} 
+ T^2 \frac{\partial}{\partial T} \left( \frac{\mu_a^*}{T} \right)_{\!\! \sigma} E_a^* \right] \right\} 
\partial_\rho u^\rho \nonumber \\
 &+& f_a^{\rm eq} \left( b_a - \frac{n_B E_a}{w} \right)
 \frac{p_a^{\mu}}{E_a^*} D_\mu \left( \frac{\mu_B}{T} \right)  \nonumber \\
 & -& f_a^{\rm eq} \, \frac{p_a^{\mu} p_a^{\nu}}{2 T E_a^*} 
 \left( D_\mu u_\nu +  D_\nu u_\mu  + \frac{2}{3} \Delta_{\mu\nu} \partial_\rho u^\rho \right) \,.
\ea
In the limit that the chemical potential goes to zero this reproduces the results of Jeon and Yaffe \cite{Jeon1996} and of Chakraborty and
Kapusta \cite{ChakrabortyKapusta2011}.

Now we subtract the right-hand side from the left-hand side and set the resulting expression to zero.  This leads to
\be
{\cal A}_a \left( \partial_{\rho} u^{\rho} \right) + {\cal B}_a^{\mu}  D_\mu \left( \frac{\mu_B}{T} \right)
-{\cal C}_a^{\mu\nu} \left( D_{\mu} u_{\nu} + D_{\nu} u_{\mu} 
+ \twoth \Delta_{\mu\nu} \partial_{\rho} u^{\rho} \right) = 0
\ee
where
\ba
{\cal A}_a &=& \frac{1}{3 T E_a^*} \left\{ |{\bf p}_a^*|^2 
 - 3 v_n^2 \left[ E_a^{*2} - T^2 \left( \frac{\partial m_a^{*2}}{\partial T^2} \right)_{\!\! \sigma} 
+ T^2 \frac{\partial}{\partial T} \left( \frac{\mu_a^*}{T} \right)_{\!\! \sigma} E_a^* \right] \right\} \nonumber \\
&+& \sum_{bcd} \frac{1}{1 + \delta_{ab}} \int d\Gamma_b^* \, d\Gamma_c^* \, d\Gamma_d^* \,
f_b^{\rm eq} \, W(a,b|c,d) \left[ A_c + A_d - A_a - A_b \right] \nonumber \\
&+& \sum_{cd} \int d\Gamma_c^* \, d\Gamma_d^* \, W(a|c,d) \left[ A_c + A_d - A_a \right] \nonumber \\
&+& \sum_{bc} \int d\Gamma_b^* \, d\Gamma_c^* \, f_b^{\rm eq} \, W(c|a,b) \left[ A_c - A_a - A_b \right]
\label{calA}
\ea
and
\ba
{\cal B}_a^{\mu} &=& \left( b_a - \frac{n_B E_a}{w} \right) \frac{p_a^{\mu}}{E_a^*} \nonumber \\
&+& \sum_{bcd} \frac{1}{1 + \delta_{ab}} \int d\Gamma_b^* \, d\Gamma_c^* \, d\Gamma_d^* \,
f_b^{\rm eq} \, W(a,b|c,d) \left[ B_c p_c^{\mu} + B_d p_d^{\mu} - B_a p_a^{\mu} - B_b p_b^{\mu} \right] \nonumber \\
&+& \sum_{cd} \int d\Gamma_c^* \, d\Gamma_d^* \, W(a|c,d) \left[ B_c p_c^{\mu} + B_d p_d^{\mu} - B_a p_a^{\mu} \right] \nonumber \\
&+& \sum_{bc} \int d\Gamma_b^* \, d\Gamma_c^* \,
f_b^{\rm eq} \, W(c|a,b) \left[ B_c p_c^{\mu} - B_a p_a^{\mu} - B_b p_b^{\mu} \right]
\label{calB}
\ea
and
\ba
{\cal C}_a^{\mu\nu} &=& \frac{p_a^{\mu} p_a^{\nu}}{2E_a^* T} \nonumber \\
&+& \sum_{bcd} \frac{1}{1 + \delta_{ab}} \int d\Gamma_b^* \, d\Gamma_c^* \, d\Gamma_d^* \, 
f_b^{\rm eq} \, W(a,b|c,d) \left[ C_c p_c^{\mu} p_c^{\nu} + C_d p_d^{\mu} p_d^{\nu}
 - C_a p_a^{\mu} p_a^{\nu} - C_b p_b^{\mu} p_b^{\nu} \right] \nonumber \\
&+& \sum_{cd} \int d\Gamma_c^* \, d\Gamma_d^* \,
W(a|c,d) \left[ C_c p_c^{\mu} p_c^{\nu} + C_d p_d^{\mu} p_d^{\nu}
 - C_a p_a^{\mu} p_a^{\nu} \right] \nonumber \\
&+& \sum_{bc} \int d\Gamma_b^* \, d\Gamma_c^* \,
f_b^{\rm eq} \, W(c|a,b) \left[ C_a p_a^{\mu} p_a^{\nu} + 
C_b p_b^{\mu} p_b^{\nu} - C_c p_c^{\mu} p_c^{\nu} \right] \,.
\label{calC}
\ea
Due to the tensorial structure of these equations the solution requires that ${\cal A}_a = 0$, ${\cal B}_a^{\mu} = 0$, and ${\cal C}_a^{\mu\nu} = 0$.  These are integral equations for the functions $A_a$, $B_a$, and $C_a$ which depend on the magnitude of the momentum ${\bf p}^*$.

\section{Landau-Lifshtiz Conditions of Fit}

The set of equations (\ref{calA})-(\ref{calC}) are integral equations for the functions $A_a$, $B_a$, and $C_a$.  Consider the equation for $A_a$.  If we have a particular solution $A_a^{\rm par}$ we can generate another solution $A_a = A_a^{\rm par} - a_E E_a - a_B b_a$ where the constant coefficients $a_E$ and $a_B$ are independent of particle type $a$.  The reason is that energy and baryon number are conserved in the collision, decay, and fusion processes.  This arbitrariness exists because of the freedom to define the local rest frame or, equivalently, the flow velocity $u^{\mu}$.   This arbitrariness is removed once one specifies the Landau-Lifshitz definition of the local rest frame, also called the condition of fit.  Requiring that $\delta T^{00} = 0$ in the local rest frame results in
\bd
a_E \sum_a \int d\Gamma_a^* E_a^2 
\left[ 1  -   \frac{ T( \partial E_a/\partial T)_{\sigma}}{E_a - \mu_a + T \left( \partial \mu_a / \partial T \right)_{\sigma} } \right] f_a^{\rm eq}
\ed
\bd
+ a_B \sum_a b_a \int d\Gamma_a^* E_a 
\left[ 1  -   \frac{ T( \partial E_a/\partial T)_{\sigma}}{E_a - \mu_a + T \left( \partial \mu_a / \partial T \right)_{\sigma} } \right] f_a^{\rm eq}
\ed
\be
= \sum_a \int d\Gamma_a^* E_a 
\left[ 1  -   \frac{ T( \partial E_a/\partial T)_{\sigma}}{E_a - \mu_a + T \left( \partial \mu_a / \partial T \right)_{\sigma} }  \right] A_a^{\rm par} \, f_a^{\rm eq} \,.
\label{energyfit}
\ee
Requiring that $\delta J_B^0 = 0$ in the local rest frame results in
\bd
a_E \sum_a b_a \int d\Gamma_a^* E_a 
\left[ 1  -   \frac{ T( \partial E_a/\partial T)_{\sigma}}{E_a - \mu_a + T \left( \partial \mu_a / \partial T \right)_{\sigma} } \right] f_a^{\rm eq}
\ed
\bd
+ a_B \sum_a b_a^2 \int d\Gamma_a^* 
\left[ 1  -   \frac{ T( \partial E_a/\partial T)_{\sigma}}{E_a - \mu_a + T \left( \partial \mu_a / \partial T \right)_{\sigma} } \right] f_a^{\rm eq}
\ed
\be
= \sum_a b_a\int d\Gamma_a^* 
\left[ 1  -   \frac{ T( \partial E_a/\partial T)_{\sigma}}{E_a - \mu_a + T \left( \partial \mu_a / \partial T \right)_{\sigma} }  \right] A_a^{\rm par} \, f_a^{\rm eq} \,.
\label{densityfit}
\ee
Let us express these equations as
\ba
a_E X_E + a_B X_B &=& Z_E \, , \nonumber \\
a_E Y_E + a_B Y_B &=& Z_B \,.
\ea
The solutions are
\ba
a_B &=& \frac{Y_E Z_E - X_E Z_B}{Y_E X_B - X_E Y_B} \, , \nonumber \\
a_E &=& \frac{X_B Z_B - Y_B Z_E}{Y_E X_B - X_E Y_B} \,.
\ea
When these are substituted into the expression (\ref{zeta}) for the bulk viscosity we get
\be
\zeta = \frac{1}{3} \sum_a \int d\Gamma_a^* \frac{|{\bf p}_a^*|^2}{E_a^*} f_a^{\rm eq} A_a -T n_B a_B - T w a_E \,.
\label{zetatemp}
\ee

First consider the case where there are no mean fields, only on-shell particles traveling in vacuum and undergoing localized collisions.  In this case $\delta f_a = \delta \tilde{f}_a$, and one finds
\ba
X_E &=& T(T^2 \chi_{TT} + 2 \mu_B T \chi_{\mu T} + \mu_B^2 \chi_{\mu\mu}) \, , \nonumber \\
X_B &=& T( T \chi_{\mu T} + \mu_B \chi_{\mu\mu}) \, , \nonumber \\
Y_E &=& T( T \chi_{\mu T} + \mu_B \chi_{\mu\mu}) \, . \nonumber \\
Y_B &=& T \chi_{\mu\mu}
\ea
The combination of $a_E$ and $a_B$ which is needed for the bulk viscosity is
\ba
T n_B a_B + T w a_E &=& v_n^2 Z_E + (v_s^2 - v_n^2 ) \mu_B Z_B \nonumber \\
&=& \sum_a \int d\Gamma_a^* \left[ v_n^2 E_a + (v_s^2 - v_n^2 ) b_a \mu_B \right] A_a^{\rm par} f_a^{\rm eq} \,.
\ea
Here $E_a = E_a^* = \sqrt{p^2 + m_a^2}$ because of the absence of mean fields.  The bulk viscosity is then
\be
\zeta = \frac{1}{3} \sum_a \int d\Gamma_a^* \left\{ \frac{|{\bf p}_a^*|^2}{E_a^*} - 3 \left[ v_n^2 E_a^* + (v_s^2 - v_n^2 ) b_a \mu_B \right] \right\}
A_a^{\rm par} f_a^{\rm eq} \,.
\ee
This is a limiting form of
\be
\zeta = \frac{1}{3} \sum_a \int d\Gamma_a^* \left[ \frac{|{\bf p}_a^*|^2}{E_a^*} + 3 v_n^2 T^2  \frac{\partial}{\partial T} \left(\frac{ E_a - \mu_a}{T} \right)_{\!\! \sigma} \right]
A_a^{\rm par} f_a^{\rm eq} \,,
\label{zetaLL}
\ee
once one recognizes Eq. (\ref{isentropic}).  This makes perfect sense because the modification of the integrand compared to Eq. (\ref{zeta}) matches the structure of the source of $A_a$ in Eq. (\ref{calA}).

It is not easy to find simple expressions for $X_E, X_B, Y_E, Y_B$ when mean fields are included, hence there are no simple expressions for $a_E$ and $a_B$.  Fortunately, the individual expressions for $a_E$ and $a_B$ are not needed to find a simple expression for the bulk viscosity.  Returning to Eq. (\ref{zeta}) we have
\be
\zeta = \frac{1}{3} \sum_a \int d\Gamma_a^* \frac{|{\bf p}_a^*|^2}{E_a^*} f_a^{\rm eq} (A_a^{\rm par} - a_E E_a - a_B b_a) \,.
\ee
Now the trick is to take a judicious linear combination of the conditions of fit.  Add $T \left( \partial \mu_{\mu_B} / \partial T \right)_{\sigma}  - \mu_B$ times (\ref{densityfit}) to (\ref{energyfit}).  This gives
\bd
a_B \sum_a b_a \int d\Gamma_a^*  \left( \frac{ \partial f_a^{\rm eq} }{\partial T} \right)_{\!\!\sigma}
 + a_E \sum_a \int d\Gamma_a^* E_a \left(  \frac{ \partial f_a^{\rm eq} }{\partial T} \right)_{\!\!\sigma}
\ed
\be
= - \sum_a \int d\Gamma_a^* f_a^{\rm eq} A_a^{\rm par}  \frac{\partial}{\partial T}
\left( \frac{E_a - \mu_a}{T} \right)_{\!\!\sigma} \,.
\label{eq:cof:aB_aE_eq1}
\ee
The coefficient of $a_B$ is just $(\partial n_B/\partial T)_{\sigma}$, and from Eq. (\ref{energyvar}) the coefficient of $a_E$ is just $(\partial \epsilon/\partial T)_{\sigma}$.  Therefore we have
\be
a_B  \left( \frac{ \partial n_B }{\partial T} \right)_{\!\!\sigma}
 + a_E \left(  \frac{ \partial \epsilon }{\partial T} \right)_{\!\!\sigma}
= - \sum_a \int d\Gamma_a^* f_a^{\rm eq} A_a^{\rm par}  \frac{\partial}{\partial T}
\left( \frac{E_a - \mu_a}{T} \right)_{\!\!\sigma} \,.
\label{eq:cof:aB_aE_eq2}
\ee

Since $v_n^2$ enters into Eq. (\ref{zetaLL}) it is useful to derive the thermodynamic relations
\be
T v_n^2 = \frac{w}{(\partial \epsilon/\partial T)_{\sigma}} = \frac{n_B}{(\partial n_B/\partial T)_{\sigma}} \,.
\label{vn2}
\ee
First, we derive the relation between the derivatives appearing in the above equations.  Using $d\epsilon = T ds + \mu_B dn_B$ and $ds = n_B d\sigma + \sigma dn_B$, we obtain
\be
\left( \frac{\partial \epsilon}{\partial T}\right)_{\!\!\sigma} = \frac{w}{n_B} \left( \frac{\partial n_B}{\partial T}\right)_{\!\!\sigma} \, .
\ee
Now for $(\partial n_B/\partial T)_{\sigma}$ we use Eq. (\ref{sigmaderivative}), the third equality of Eq. (\ref{isentropic}), and the first equality of Eq. (\ref{speeds}) to obtain
\be
T \left( \frac{\partial n_B}{\partial T}\right)_{\!\!\sigma} = \frac{n_B}{v_n^2} \,.
\ee
Together with the previous equation we obtain the desired result (\ref{vn2}).  Using these results in Eq. (\ref{eq:cof:aB_aE_eq2}) we have
\be
T n_B a_B + T w a_E = - v_n^2 T^2 \sum_a \int d\Gamma_a^* f_a^{\rm eq} A_a^{\rm par}  \frac{\partial}{\partial T}
\left( \frac{E_a - \mu_a}{T} \right)_{\!\!\sigma} \,.
\label{eq:cof:aB_aE_eq3}
\ee
Making this substitution in Eq. (\ref{zetatemp}) we obtain the expression (\ref{zetaLL}).

A similar arbitrariness arises in Eq. (\ref{calB}).  Due to energy-momentum conservation, if we have a particular solution $B_a^{\rm par}$ we can generate another solution as $B_a = B_a^{\rm par} - b$, where $b$ is a constant independent of particle species $a$.  This freedom is resolved by the Landau-Lifshitz condition of fit which requires that $\delta T^{0j} = 0$ in the local rest frame.  Starting with expression (\ref{dspace-time}) we have
\be
\delta T^{0j} = \sum_a \int d\Gamma_a^* \frac{p_a^{*j}}{E_a^*} E_a \left[ - \left( B_a^{\rm par} - b \right) p_a^{*i} D_i \left(\frac{\mu_B}{T} \right) \right] f_a^{\rm eq} \,.
\ee
Factoring out the spatial derivative, and making use of the momentum space isotropy, we require that
\be
b \sum_a \int d\Gamma_a^* \frac{|p_a^*|^2}{E_a^*} E_a f_a^{\rm eq} =
\int d\Gamma_a^* \frac{|p_a^*|^2}{E_a^*} E_a B_a^{\rm par} f_a^{\rm eq} \,.
\ee
The integral multiplying $b$ is just $3Tw$ so that
\be
b  = \frac{1}{3Tw} \int d\Gamma_a^* \frac{|p_a^*|^2}{E_a^*} E_a B_a^{\rm par} f_a^{\rm eq} \,.
\ee
Substitution into expression (\ref{lambda}) gives 
\be
\lambda = \frac{1}{3} \left( \frac{w}{n_B T} \right)^2 \sum_a \int d\Gamma_a^* \frac{|{\bf p}_a^*|^2}{E_a^*} 
\left( b_a - \frac{n_B E_a}{w} \right) B_a^{\rm par} f_a^{\rm eq} \,.
\label{lambdaLL}
\ee

There is no ambiguity in the solution to Eq. (\ref{calC}) for $C_a$, so the expression for the shear viscosity (\ref{eta}) is unchanged.

\section{Relaxation-Time Approximation}

At this point, it is convenient to derive the relaxation time approximation formulas for the shear and bulk viscosities and thermal conductivity.  We start with the Boltzmann equation with the Chapman-Enskog expansion:
\be
 \frac{d f_a^{\rm eq}}{dt} = {\cal C}_a \,.
 \label{eq:RTA:BTE_general}
\ee
The left-hand side of Eq. \ref{eq:RTA:BTE_general} is given by Eq. \ref{eq:chapman:BTE_LHS6} while ${\cal C}_a$ can be found in Eq. \ref{eq:collisionintegral:collision_full2}.  In the energy-dependent relaxation time approximation \cite{ChakrabortyKapusta2011}, we assume particle species $a$ is out of equilibrium ($\phi_a \ne 0$) while all other particle species are in equilibrium ($\phi_b = \phi_c = \phi_d = 0$).  Using Eq. \ref{eq:collisionintegral:collision_full2}, the collision integral ${\cal C}_a$ greatly simplifies, and the Boltzmann equation becomes
\be
 \frac{d f_a^{\rm eq}}{dt} = {\cal C}_a =  - \frac{f_a^{\rm eq} \phi_a}{\tau_a} 
 \label{eq:RTA:BTE_RTA}
\ee
where the relaxation time $\tau_a(E_a^*)$ for species $a$ is given by
\ba
 \frac{1}{\tau_a(E_a^*)} 
 &=&  \sum_{bcd} \frac{1}{1+\delta_{ab}}
 \int d\Gamma_b^* \, d\Gamma_c^* \, d\Gamma_d^* \,
 f_b^{\rm eq} W(a,b|c,d)  \nonumber \\
 &+&  \sum_{cd} \int d\Gamma_c^* \, d\Gamma_d^* \, W(a|c,d) 
 \nonumber \\
 &+& \sum_{bc} \int d\Gamma_b^* \, d\Gamma_c^* \,
 f_b^{\rm eq} W(c|a,b)  \,.
 \label{eq:RTA:relaxationtime}
\ea

Next, we replace the left-hand side of Eq. \ref{eq:RTA:BTE_RTA} using Eq. \ref{eq:chapman:BTE_LHS6}. Into the right-hand side, we substitute $\phi_a$ using Eq. \ref{eq:derivingphi:phi}.  Then we equate terms on the left- and right-hand sides by matching tensor structures, and we obtain
particular solutions for the functions $A_a^{}$, $B_a$, and $C_a$ from $\phi_a$:
\be
  A_a^{\rm par} = \frac{\tau_a}{3T} \left[ \frac{ |{\bf p}_a^*|^2 }{ E_a^*} 
 +  3 v_n^2 T^2 \frac{\partial}{\partial T} \left(\frac{ E_a - \mu_a}{T} \right)_{\!\! \sigma} \right] \, ,
 \label{eq:RTA:Apar}
\ee
\be
  B_a^{\rm par} = \frac{\tau_a}{E_a^*} \left( b_a - \frac{n_B E_a}{w} \right) \, ,
  \label{eq:RTA:Bpar}
\ee 
\be
 C_a^{\rm par} = \frac{\tau_a}{2TE_a^*} \,.
 \label{eq:RTA:Cpar}
\ee 
Finally, we substitute Eqs. \ref{eq:RTA:Apar}-\ref{eq:RTA:Cpar} into Eqs. \ref{eta}, \ref{zetaLL}, and \ref{lambdaLL} and obtain the desired relaxation time formulas:
\be
 \eta = \frac{1}{15T} \sum_a \int d\Gamma_a^* \frac{|{\bf p}_a^*|^4}{E_a^{*2}}
 \tau_a(E_a^*) f_a^{\rm eq} \, ,
 \label{etaRTA}
\ee
\be
 \zeta = \frac{1}{9T} \sum_a \int d\Gamma_a^* 
 \frac{\tau_a(E_a^*)}{E_a^{*2}}
 \left[ |{\bf p}_a^*|^2 + 3 v_n^2 T^2 E_a^* \frac{\partial}{\partial T} \left(\frac{ E_a - \mu_a}{T} \right)_{\!\! \sigma} \right]^2
 f_a^{\rm eq} \, ,
 \label{zetaRTA}
\ee
\be
 \lambda = \frac{1}{3} \left( \frac{w}{n_B T} \right)^2 \sum_a 
 \int d\Gamma_a^* \frac{|{\bf p}_a^*|^2}{E_a^{*2}} 
 \tau_a(E_a^*)
 \left( b_a - \frac{n_B E_a}{w} \right)^2  f_a^{\rm eq} \,.
 \label{lambdaRTA}
\ee

A few observations are in order.  First, the transport coefficients computed with Eqs. \ref{etaRTA}-\ref{lambdaRTA} are strictly non-negative, as they must be.  Second, this non-negativity is ensured by the squares in the integrands which came from enforcing the Landau-Lifshitz conditions of fit.  (Recall the derivation of Eqs. \ref{zetaLL} and \ref{lambdaLL}.)  This shows that it is absolutely vital that the Landau-Lifshitz conditions are carefully enforced in order to obtain the correct results.  A third point is that Eqs. \ref{etaRTA} and \ref{zetaRTA} are obvious generalizations of the formulas obtained in previous works \cite{ChakrabortyKapusta2011,Jeon1996} to finite baryon chemical potential.  The crucial insight is that entropy per baryon ($\sigma = s/n_B$) is conserved in zeroth-order (ideal) hydrodynamics, so that variable that must be held fixed when deriving the variations from equilibrium.

\section{Conclusion}

In this paper, we developed a flexible relativistic quasiparticle theory of transport coefficients in hot and dense hadronic matter.  A major goal was the simultaneous inclusion of temperature- and baryon chemical potential-dependent quasiparticle masses with scalar and vector mean fields, all in a thermodynamically self-consistent way.  Classical statistics were used throughout to simplify the presentation, although complete results with quantum statistics are given in the appendix.  From the dispersion relations for the quasiparticles, we derived the Boltzmann equation and then the transport coefficients using the Chapman-Enskog expansion.  Next, we derived compact analytic expressions for the shear and bulk viscosities and thermal conductivity.  These formulas can be used with the relaxation time approximation; alternatively, we have provided integral equations which may be solved for greater accuracy.
We have shown that the transport coefficients are non-negative in the relaxation time approximation (as they must be) which is a direct consequence of carefully enforcing the Landau-Lifshitz conditions of fit.

We also showed that previous bulk viscosity formulas (derived assuming zero baryon chemical potential) generalize straightforwardly to finite baryon chemical potential
if one recalls that entropy per baryon is conserved in ideal hydrodynamics.  This was the crucial detail that allowed us to compute the variations from equilibrium and use them to derive the bulk viscosity and thermal conductivity formulas.

It is a trivial matter to include a variety of scalar and vector fields, that is simply a matter of book-keeping.  The same is true of additional conserved charges beyond baryon number.  In future work we will study specific hadronic models, including numerical solutions, along the lines of Ref. \cite{ChakrabortyKapusta2011}.

\section*{Acknowledgments}

This work was supported by the US Department of Energy (DOE) under Grant No. DE-FG02-87ER40328.

\section*{Appendix}

This appendix has two goals.  The first is to summarize the important results derived in the main body of the manuscript.  The second is to include the effects of quantum statistics.  All results presented here include quantum statistics.  The limit of classical statistics is attained when $|f_a| \ll 1$.

Departures from local kinetic and chemical equilibrium for particle species $a$ are expressed in terms of the function $\phi_a$ as 
\be
f_a = f_a^{\rm eq} \left( 1 + \phi_a \right) \,.
\ee
We let $\delta f_a$ represent the deviation expressed in terms of the equilibrium energy $E_a^0$ while $\delta \tilde{f}_a$ represents the deviation expressed in terms of the total nonequilibrium energy $E_a$; it is the latter which is conserved in local collisions and the one relevant for transport coefficients.  The deviations to each other by
\be
\delta f_a = \delta \tilde{f}_a + \left( \frac{\partial f_a^{\rm eq}}{\partial E_a} \right)_{T^0 ,\, \mu_B^0} \delta E_a
= \delta \tilde{f}_a - \frac{\delta E_a}{T} f_a^{\rm eq} \left( 1 + d_a  f_a^{\rm eq} \right) \,.
\label{ftildefQ}
\ee
Here the notation is $d_a = (-1)^{2s_a}$.  We need to relate the variations in $T$ and $\mu_B$ to the variation $\delta \tilde{f}_a$.  The latter variation is done at fixed $E_a$ and is
\be
\delta \tilde{f}_a = f_a^{\rm eq} \left[ E_a - \mu_a + T \left( \frac{\partial \mu_a}{\partial T} \right)_{\!\! \sigma} \right]
\left( 1 + d_a  f_a^{\rm eq} \right) \frac{\delta T}{T^2}
\label{dtildefQ}
\ee
Here in what follows, the derivative is carried out at fixed entropy per baryon $\sigma$.  The factor from Eq. (\ref{ftildefQ}) which needs to be rewritten is
\be
\frac{\delta E_a}{T} f_a^{\rm eq} \left( 1 + d_a  f_a^{\rm eq} \right)
= \left[ \frac{ T( \partial E_a/\partial T)_{\sigma}}{E_a - \mu_a + T \left( \partial \mu_a / \partial T \right)_{\sigma} } \right] \delta \tilde{f}_a \,.
\ee
In terms of $\delta \tilde{f}_a$ the deviations in the energy-momentum tensor and baryon current are as follows.
\be
\delta T^{ij} = \sum_a \int d\Gamma_a^* \frac{p_a^{*i} p_a^{*j}}{E_a^*} \delta \tilde{f}_a
\label{dTijQ}
\ee
\be
\delta T^{0j} = \sum_a \int d\Gamma_a^* \frac{p_a^{*j}}{E_a^*} E_a \delta \tilde{f}_a \,.
\label{dspace-timeQ} 
\ee
\be
\delta T^{00} = \sum_a \int d\Gamma_a^* E_a
 \left\{ 1  -   \frac{ T( \partial E_a/\partial T)_{\sigma}}{E_a - \mu_a + T \left( \partial \mu_a / \partial T \right)_{\sigma} }\right\} \delta \tilde{f}_a \,. 
\ee
\be
\delta J_B^i = \sum_a b_a \int d\Gamma_a^* \frac{p_a^{*i}}{E_a^*} \delta \tilde{f}_a \,.
\ee
\be
\delta J_B^0 = \sum_a b_a \int d\Gamma_a^* 
\left\{ 1  -   \frac{ T( \partial E_a/\partial T)_{\sigma}}{E_a - \mu_a + T \left( \partial \mu_a / \partial T \right)_{\sigma} }  \right\} \delta \tilde{f}_a 
\ee

The collision term on the right side of the Boltzmann equation reads
\bd
 \mathcal{C}_a =
\sum_{bcd} \frac{1}{1 + \delta_{ab}} \int d\Gamma_b^* \, 
d\Gamma_c^* \, d\Gamma_d^* \, W(a,b|c,d)
\ed
\bd 
\times \Big\{ 
f_c f_d \left( 1 + d_a f_a \right) \left( 1 + d_b f_b \right)
- f_a f_b \left( 1 + d_c f_c \right) \left( 1 + d_d f_d \right) 
\Big\}
\ed
\bd
+ \sum_{cd} \int d\Gamma_c^* \, d\Gamma_d^* \, W(a|c,d) 
\ed
\bd
\times \Big\{ f_c f_d \left( 1 + d_a f_a \right)  
 - f_a \left( 1 + d_c f_c \right) \left( 1 + d_d f_d \right)
\Big\}
\ed
\bd
+ \sum_{bc} \int d\Gamma_b^* \, d\Gamma_c^* \, W(c|a,b) 
\ed
\be
\times
\Big\{ f_c
\left( 1 + d_a f_a \right) \left( 1 + d_b f_b \right)  
 - f_a f_b \left( 1 + d_c f_c \right) 
\Big\} \, .
\ee
This expression explicitly includes $2 \leftrightarrow 2$ and $2 \leftrightarrow 1$ reactions.  Higher order reactions are included in an obvious way.

We now consider small departures from equilibrium, meaning that we keep terms only linear in the $\phi_a$.  We use chemical equilibrium; for example, $a + b \leftrightarrow c + d$ gives 
\bd
f_c^{\rm eq} f_d^{\rm eq} \left( 1 + d_a f_a^{\rm eq} \right) \left( 1 + d_b f_b^{\rm eq} \right) = 
f_a^{\rm eq} f_b^{\rm eq} \left( 1 + d_c f_c^{\rm eq} \right) \left( 1 + d_d f_d^{\rm eq} \right)  \,.
\ed
Then the collision integral becomes
\ba
 \mathcal{C}_a 
 &=& \sum_{bcd} \frac{1}{1+\delta_{ab}}
 \int d\Gamma_b^* \, d\Gamma_c^* \, d\Gamma_d^* \,W(a,b|c,d) \nonumber \\
& \times & \Big\{ f_a^{\rm eq} f_b^{\rm eq} \left[ \left( 1 + d_d f_d^{\rm eq} \right) \phi_c + \left( 1 + d_c f_c^{\rm eq} \right) \phi_d \right]
 - f_c^{\rm eq} f_d^{\rm eq} \left[ \left( 1 + d_b f_b^{\rm eq} \right) \phi_a + \left( 1 + d_a f_a^{\rm eq} \right) \phi_b \right] \Big\} \nonumber \\
 &+& \sum_{cd} \int d\Gamma_c^* \, d\Gamma_d^* \, W(a|c,d) \Big\{ f_a^{\rm eq}  \left[ \left( 1 + d_d f_d^{\rm eq} \right) \phi_c 
+ \left( 1 + d_c f_c^{\rm eq} \right) \phi_d \right] - f_c^{\rm eq} f_d^{\rm eq} \phi_a \Big\}
 \nonumber \\
 &+& \sum_{bc} \int d\Gamma_b^* \, d\Gamma_c^* \, W(c|a,b) \Big\{ - f_c^{\rm eq}  \left[ \left( 1 + d_b f_b^{\rm eq} \right) \phi_a 
+ \left( 1 + d_a f_a^{\rm eq} \right) \phi_b \right] + f_a^{\rm eq} f_b^{\rm eq} \phi_c \Big\} \,.
 \label{eq:collisionintegral:collision_full2Q}
\ea

The left-hand side of the Boltzmann equation is computed using the local equilibrium form of the distribution function.  One form of the left-hand side (in the local rest frame) is
\ba
\frac{d f_a^{\rm eq}}{dt} &=&  f_a^{\rm eq} \left( 1 + d_a f_a^{\rm eq} \right) \left[  \frac{ |{\bf p}_a^*|^2 }{3 T E_a^*} 
 +  v_n^2 T \frac{\partial}{\partial T} \left(\frac{ E_a - \mu_a}{T} \right)_{\!\! \sigma} \right] \partial_\rho u^\rho \nonumber \\
 &+& f_a^{\rm eq} \left( 1 + d_a f_a^{\rm eq} \right) \left( b_a - \frac{n_B E_a}{w} \right)
 \frac{p_a^{\mu}}{E_a^*} D_\mu \left( \frac{\mu_B}{T} \right)  \nonumber \\
 & -& f_a^{\rm eq} \left( 1 + d_a f_a^{\rm eq} \right) \, \frac{p_a^{\mu} p_a^{\nu}}{2 T E_a^*} 
 \left( D_\mu u_\nu +  D_\nu u_\mu  + \frac{2}{3} \Delta_{\mu\nu} \partial_\rho u^\rho \right) \,.
\label{eq:chapman:BTE_LHS6Q}
\ea

Now we subtract the right-hand side from the left-hand side and set the resulting expression to zero.  This leads to
\be
{\cal A}_a \left( \partial_{\rho} u^{\rho} \right) + {\cal B}_a^{\mu}  D_\mu \left( \frac{\mu_B}{T} \right)
-{\cal C}_a^{\mu\nu} \left( D_{\mu} u_{\nu} + D_{\nu} u_{\mu} 
+ \twoth \Delta_{\mu\nu} \partial_{\rho} u^{\rho} \right) = 0
\ee
where
\ba
{\cal A}_a &=&  \left[  \frac{ |{\bf p}_a^*|^2 }{3 T E_a^*} 
 +  v_n^2 T \frac{\partial}{\partial T} \left(\frac{ E_a - \mu_a}{T} \right)_{\!\! \sigma} \right] f_a^{\rm eq} \left( 1 + d_a f_a^{\rm eq} \right) \nonumber \\
&+& \sum_{bcd} \frac{1}{1+\delta_{ab}} \int d\Gamma_b^* \, d\Gamma_c^* \, d\Gamma_d^* \,W(a,b|c,d) \nonumber \\
& \times & \Big\{ f_a^{\rm eq} f_b^{\rm eq} \left[ \left( 1 + d_d f_d^{\rm eq} \right) A_c + \left( 1 + d_c f_c^{\rm eq} \right) A_d \right] \nonumber \\
&-& f_c^{\rm eq} f_d^{\rm eq} \left[ \left( 1 + d_b f_b^{\rm eq} \right) A_a + \left( 1 + d_a f_a^{\rm eq} \right) A_b \right] \Big\} \nonumber \\
 &+& \sum_{cd} \int d\Gamma_c^* \, d\Gamma_d^* \, W(a|c,d) \Big\{ f_a^{\rm eq}  \left[ \left( 1 + d_d f_d^{\rm eq} \right) A_c 
+ \left( 1 + d_c f_c^{\rm eq} \right) A_d \right] - f_c^{\rm eq} f_d^{\rm eq} A_a \Big\} \nonumber \\
 &+& \sum_{bc} \int d\Gamma_b^* \, d\Gamma_c^* \, W(c|a,b) \Big\{ - f_c^{\rm eq}  \left[ \left( 1 + d_b f_b^{\rm eq} \right) A_a 
+ \left( 1 + d_a f_a^{\rm eq} \right) A_b \right] + f_a^{\rm eq} f_b^{\rm eq} A_c \Big\} \,, \nonumber \\
\label{calAQ}
\ea
\ba
{\cal B}_a^{\mu} &=& \left( b_a - \frac{n_B E_a}{w} \right) \frac{p_a^{\mu}}{E_a^*} f_a^{\rm eq} \left( 1 + d_a f_a^{\rm eq} \right)  \nonumber \\
&+& \sum_{bcd} \frac{1}{1 + \delta_{ab}} \int d\Gamma_b^* \, d\Gamma_c^* \, d\Gamma_d^* \, W(a,b|c,d) \nonumber \\
& \times & \Big\{ f_a^{\rm eq} f_b^{\rm eq} \left[ \left( 1 + d_d f_d^{\rm eq} \right) B_c p_c^{\mu} + \left( 1 + d_c f_c^{\rm eq} \right) B_d p_d^{\mu} \right] \nonumber \\
&-& f_c^{\rm eq} f_d^{\rm eq} \left[ \left( 1 + d_b f_b^{\rm eq} \right) B_a p_a^{\mu} + \left( 1 + d_a f_a^{\rm eq} \right) B_b p_b^{\mu} \right] \Big\} \nonumber \\
 &+& \sum_{cd} \int d\Gamma_c^* \, d\Gamma_d^* \, W(a|c,d) \Big\{ f_a^{\rm eq}  \left[ \left( 1 + d_d f_d^{\rm eq} \right) B_c p_c^{\mu} 
+ \left( 1 + d_c f_c^{\rm eq} \right) B_d p_d^{\mu} \right] - f_c^{\rm eq} f_d^{\rm eq} B_a p_a^{\mu} \Big\} \nonumber \\
 &+& \sum_{bc} \int d\Gamma_b^* \, d\Gamma_c^* \, W(c|a,b) \Big\{ - f_c^{\rm eq}  \left[ \left( 1 + d_b f_b^{\rm eq} \right) B_a p_a^{\mu} 
+ \left( 1 + d_a f_a^{\rm eq} \right) B_b p_b^{\mu} \right] + f_a^{\rm eq} f_b^{\rm eq} B_c p_c^{\mu} \Big\} \,, \nonumber \\ 
\label{calBQ}
\ea
\ba
{\cal C}_a^{\mu\nu} &=& \frac{p_a^{\mu} p_a^{\nu}}{2E_a^* T}  f_a^{\rm eq} \left( 1 + d_a f_a^{\rm eq} \right) \nonumber \\
&+& \sum_{bcd} \frac{1}{1 + \delta_{ab}} \int d\Gamma_b^* \, d\Gamma_c^* \, d\Gamma_d^* \, W(a,b|c,d) \nonumber \\
& \times & \Big\{ f_a^{\rm eq} f_b^{\rm eq} \left[ \left( 1 + d_d f_d^{\rm eq} \right) C_c p_c^{\mu}p_c^{\nu} + 
\left( 1 + d_c f_c^{\rm eq} \right) C_d p_d^{\mu}p_d^{\nu} \right] \nonumber \\
&-& f_c^{\rm eq} f_d^{\rm eq} \left[ \left( 1 + d_b f_b^{\rm eq} \right) C_a p_a^{\mu}p_a^{\nu} + \left( 1 + d_a f_a^{\rm eq} \right) C_b p_b^{\mu}p_b^{\nu} \right] \Big\} \nonumber \\
 &+& \sum_{cd} \int d\Gamma_c^* \, d\Gamma_d^* \, W(a|c,d) \Big\{ f_a^{\rm eq}  \left[ \left( 1 + d_d f_d^{\rm eq} \right) C_c p_c^{\mu}p_c^{\nu}
+ \left( 1 + d_c f_c^{\rm eq} \right) C_d p_d^{\mu}p_d^{\nu} \right] - f_c^{\rm eq} f_d^{\rm eq} C_a p_a^{\mu}p_a^{\mu} \Big\} \nonumber \\
 &+& \sum_{bc} \int d\Gamma_b^* \, d\Gamma_c^* \, W(c|a,b) \Big\{ - f_c^{\rm eq}  \left[ \left( 1 + d_b f_b^{\rm eq} \right) C_a p_a^{\mu}p_a^{\nu} 
+ \left( 1 + d_a f_a^{\rm eq} \right) C_b p_b^{\mu}p_b^{\nu} \right] + f_a^{\rm eq} f_b^{\rm eq} C_c p_c^{\mu}p_c^{\nu} \Big\} \,. \nonumber \\ 
\label{calCQ}
\ea
Due to the tensorial structure of these equations the solution requires that ${\cal A}_a = 0$, ${\cal B}_a^{\mu} = 0$, and ${\cal C}_a^{\mu\nu} = 0$.  These are integral equations for the functions $A_a$, $B_a$, and $C_a$ which depend on the magnitude of the momentum ${\bf p}^*$.

The solutions for $A_a$ and $B_a$ are not unique.  It is necessary to specify whether $u^{\mu}$ represents the flow of energy (Landau-Lifshitz) or baryon number (Eckart).  We enforce the Landau-Lifshitz condition, sometimes known as the condition of fit, using any particular solutions.  The results for the transport coefficients are
\be
\zeta = \frac{1}{3} \sum_a \int d\Gamma_a^* \left[ \frac{|{\bf p}_a^*|^2}{E_a^*} + 3 v_n^2 T^2  \frac{\partial}{\partial T} \left(\frac{ E_a - \mu_a}{T} \right)_{\!\! \sigma} \right]
A_a^{\rm par} f_a^{\rm eq} \,,
\label{zetaLLQ}
\ee
\be
\lambda = \frac{1}{3} \left( \frac{w}{n_B T} \right)^2 \sum_a \int d\Gamma_a^* \frac{|{\bf p}_a^*|^2}{E_a^*} 
\left( b_a - \frac{n_B E_a}{w} \right) B_a^{\rm par} f_a^{\rm eq} \,,
\label{lambdaLLQ}
\ee
\be
\eta = \frac{2}{15} \sum_a \int d\Gamma_a^* \frac{|{\bf p}_a^*|^4}{E_a^*}  C_a^{\rm par} f_a^{\rm eq} \,.
\label{etaQ}
\ee
The particular solutions need not even satisfy the Boltzmann equation to satisfy the condition of fit.

A common approximation is the energy-dependent relaxation time approximation.  It assumes that only one $\phi_a$ is nonzero and the others vanish.  Then the Boltzmann equation is approximated by
\be
 \frac{d f_a^{\rm eq}}{dt} = {\cal C}_a =  - \frac{f_a^{\rm eq} \phi_a}{\tau_a} \,,
 \label{eq:RTA:BTE_RTAQ}
\ee
where the relaxation time $\tau_a(E_a^*)$ for species $a$ is given by
\ba
 \frac{1+d_a f_a^{\rm eq}}{\tau_a(E_a^*)}  &=&  \sum_{bcd} \frac{1}{1+\delta_{ab}}
 \int d\Gamma_b^* \, d\Gamma_c^* \, d\Gamma_d^* \,  W(a,b|c,d) f_b^{\rm eq}  \left( 1 + d_c f_c^{\rm eq} \right) \left( 1 + d_d f_d^{\rm eq} \right)  \nonumber \\
 &+&  \sum_{cd} \int d\Gamma_c^* \, d\Gamma_d^* \, W(a|c,d) \left( 1 + d_c f_c^{\rm eq} \right) \left( 1 + d_d f_d^{\rm eq} \right) \nonumber \\
 &+& \sum_{bc} \int d\Gamma_b^* \, d\Gamma_c^* \, W(c|a,b)  f_b^{\rm eq} \left( 1 + d_c f_c^{\rm eq} \right) \,.
 \label{eq:RTA:relaxationtimeQ}
\ea
The particular solutions are
\be
  A_a^{\rm par} = \frac{\tau_a}{3T} \left[ \frac{ |{\bf p}_a^*|^2 }{ E_a^*} 
 +  3 v_n^2 T^2 \frac{\partial}{\partial T} \left(\frac{ E_a - \mu_a}{T} \right)_{\!\! \sigma} \right] \left( 1 + d_a f_a^{\rm eq} \right) \,
 \label{eq:RTA:AparQ}
\ee
\be
  B_a^{\rm par} = \frac{\tau_a}{E_a^*} \left( b_a - \frac{n_B E_a}{w} \right) \left( 1 + d_a f_a^{\rm eq} \right) \,,
  \label{eq:RTA:BparQ}
\ee 
\be
 C_a^{\rm par} = \frac{\tau_a}{2TE_a^*} \left( 1 + d_a f_a^{\rm eq} \right) \,.
 \label{eq:RTA:CparQ}
\ee 
Substitution gives the transport coefficients
\be
 \eta = \frac{1}{15T} \sum_a \int d\Gamma_a^* \frac{|{\bf p}_a^*|^4}{E_a^{*2}}
 \tau_a(E_a^*) f_a^{\rm eq} \left( 1 + d_a f_a^{\rm eq} \right) \,,
 \label{etaRTAQ}
\ee
\be
 \zeta = \frac{1}{9T} \sum_a \int d\Gamma_a^* 
 \frac{\tau_a(E_a^*)}{E_a^{*2}}
 \left[ |{\bf p}_a^*|^2 + 3 v_n^2 T^2 E_a^* \frac{\partial}{\partial T} \left(\frac{ E_a - \mu_a}{T} \right)_{\!\! \sigma} \right]^2
 f_a^{\rm eq} \left( 1 + d_a f_a^{\rm eq} \right) \,,
 \label{zetaRTAQ}
\ee
\be
 \lambda = \frac{1}{3} \left( \frac{w}{n_B T} \right)^2 \sum_a 
 \int d\Gamma_a^* \frac{|{\bf p}_a^*|^2}{E_a^{*2}} 
 \tau_a(E_a^*)
 \left( b_a - \frac{n_B E_a}{w} \right)^2  f_a^{\rm eq} \left( 1 + d_a f_a^{\rm eq} \right) \,.
 \label{lambdaRTAQ}
\ee
These are clearly positive-definite.

\end{document}